\documentclass[unsortedaddress,aps,preprint]{revtex4}
\usepackage{graphicx}
\usepackage{dcolumn}
\usepackage{amsmath}
\usepackage{amssymb}
\usepackage{amsfonts}
\usepackage{bm}
\usepackage{color}
\usepackage[normalem]{ulem}
\usepackage{hyperref}
\hypersetup{colorlinks=true,linkcolor=blue,urlcolor=magenta}
\newcommand{\be}{\begin{equation}}
\newcommand{\ee}{\end{equation}}
\newcommand{\ba}{\begin{eqnarray}}
\newcommand{\ea}{\end{eqnarray}}

\begin{document}
\title{Condensation and crystal nucleation in a lattice gas with a realistic phase diagram}
\author{Santi Prestipino$^1$\footnote{Email: {\tt sprestipino@unime.it}} and Gabriele Costa$^1$}
\affiliation{$^1$Dipartimento di Scienze Matematiche ed Informatiche, Scienze Fisiche e Scienze della Terra, Universit\`a degli Studi di Messina, Viale F. Stagno d'Alcontres 31, 98166 Messina, Italy}

\begin{abstract}We reconsider model II of [{\em J. Chem. Phys.} {\bf 1968}, {\em 49}, 1778--1783], a two-dimensional lattice-gas system featuring a crystalline phase and two distinct fluid phases (liquid and vapor). In this system, a particle prevents other particles from occupying sites up to third neighbors on the square lattice, while attracting (with decreasing strength) particles sitting at fourth- or fifth-neighbor sites. To make the model more realistic, we assume a finite repulsion at third-neighbor distance, with the result that a second crystalline phase appears at higher pressures. However, the similarity with real-world substances is only partial: on closer inspection the alleged liquid-vapor transition turns out to be a continuous (albeit sharp) crossover, even near the putative triple point. Closer to the standard picture is instead the freezing transition, as we show by computing the free-energy barrier to crystal nucleation from the "liquid".
\end{abstract}
\maketitle
\section{Introduction}

An age-old question in statistical physics is to what extent particles residing on the nodes of a regular lattice can reproduce the emergent properties of a continuous many-body system, like e.g. those encoded in the phase diagram~\cite{runnels1,orban,hall,poland,prestipino1,prestipino2}. Leaving aside systems like the Potts lattice gas~\cite{shih,conner} or the mixture of hard hexagons and points \cite{vanDuijneveldt}, whose phase diagrams also recall that of a simple fluid, we restrict our discussion to one-component lattice gases. This class of models has been recently revitalized by a series of computational studies~\cite{panagiotopoulos,fernandes,ramola,nath1,nath2,thewes,jaleel1,jaleel2} aimed at establishing how the order-disorder transition of hard-core lattice particles depends on the range of the forbidden region and whether the order of the phase transition can be anticipated by symmetry considerations. Systems of asymmetric hard-core particles have also been investigated~\cite{dickman,kundu,mandal}, but here we focus on isotropic interactions. As a general rule, the more extended is the range of exclusion around a particle, the more ``conventional'' is the melting behavior. While the solid-liquid transition is dominated by strong short-range repulsive forces, a sufficiently long-ranged attraction is needed to promote a liquid-vapor transition~\cite{prestipino1,bolhuis}.

Without wanting to make an exhaustive analysis of the problem, we fix our attention at model II of \cite{orban}, probably the simplest instance of a lattice-gas system with three phases (solid, liquid, and vapor). With interactions extending up to fifth-neighbor sites on the square lattice, this model gets the miracle of a phase diagram of standard type with a limited number of ingredients. Furthermore, the model can be refined (see Section 2) in such a way as to induce another crystalline phase at higher pressures, thus making it even more appealing. All these conclusions are drawn from transfer-matrix calculations, which however are only feasible for lattice strips of relatively small width. Hence, a supplement of analysis is needed to conclude that the indications of the transfer matrix are genuine, i.e., reflect the transition behavior of a simple fluid. To this aim, we run grand-canonical Monte Carlo simulations across the purported liquid-vapor coexistence line, not far away from the putative triple point, in order to see whether the peak of compressibility grows with lattice size as dictated by the theory of finite-size scaling. As for the freezing transition, the litmus test will be to find a strong hysteresis from either side of the coexistence line and/or hints that the transition is thermally activated.

The paper is organized as follows. In the next Section we describe the model and the method employed. In Section 3 we present and discuss our transfer-matrix and simulation data (additional information that would be too cumbersome to be included in the body of the paper is put in two appendixes). Conclusions follow in Section 4.

\section{Model and method}

\begin{figure}
\begin{center}
\includegraphics[width=12cm]{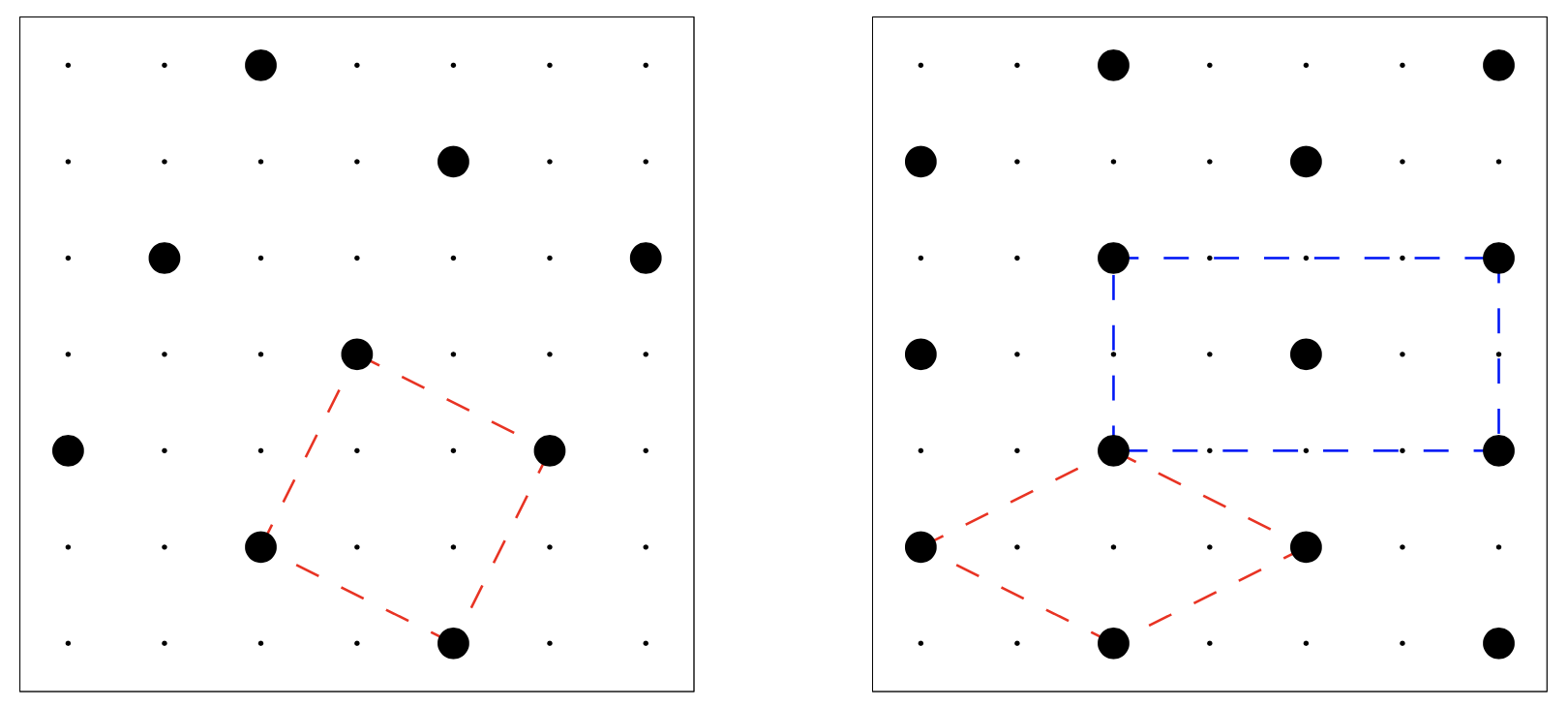}
\end{center}
\caption{The two stable crystals of the MOVB model (particular). Left: square crystal. The distance between two neighboring particles is $r_4=\sqrt{5}a$. A primitive unit cell is shown in red. Right: centered-rectangular crystal. Each particle in this crystal has two neighboring particles at distance $r_3=2a$ and other four particles at distance $r_4$. A primitive unit cell (red) and a non-primitive cell (blue) are shown.}
\end{figure}

We study a lattice-gas model on the square lattice, with a spherically-symmetric interaction extending up to fifth-neighbor sites. Calling $c_i$ (0 or 1) the occupation number of the $i$th site ($i=1,\ldots,N_s$) and $r_n$ the $n$th-neighbor distance, the system Hamiltonian reads $H=\sum_{i<j}u(r_{ij})c_ic_j$ with
\be
u(r)=\left\{
\begin{array}{ll}
+\infty & \,\,,\,{\rm for}\,\,\,r=r_1\,\,{\rm or}\,\,r_2\\
1.3\epsilon & \,\,,\,{\rm for}\,\,\,r=r_3 \\
-1.2\epsilon & \,\,,\,{\rm for}\,\,\,r=r_4 \\
-\epsilon & \,\,,\,{\rm for}\,\,\,r=r_5
\end{array}
\right.
\label{eq1}
\ee
In the above equation, $\epsilon>0$ is an arbitrary energy unit. This model departs only slightly from model II of Orban, Van Craen, and Bellemans~\cite{orban} (hence the name ``modified OVB'' or MOVB model), the only difference lying in the extension of the core: the originally infinite repulsion at third-neighbor distance is replaced in the MOVB model with a finite-strength repulsion. Accordingly, denoting with $a$ the lattice step, the maximum value of the particle-number density changes from $(1/5)a^{-2}$ (for the square crystal in the left panel of Fig.\,1) to $(1/4)a^{-2}$. Interestingly, the square crystal with density $(1/4)a^{-2}$ is not the unique configuration with this density, since a one-step shift of a line of particles would keep the density unchanged (while giving rise to a different system configuration). Among the infinite number of close-packed configurations, the one with minimum energy is the centered-rectangular (c-ret) crystal represented in the right panel of Fig.\,1: only this crystal ensures an optimum of four particles at a distance $r_4$ apart from any given particle. For lower pressures, however, the stable solid at zero temperature ($T=0$) is the square crystal in Fig.\,1 left panel, which holds the minimum energy content among {\em all} configurations. To locate the transition between the two crystals, one simply observes that for $T=0$ the energy and particle number of the square crystal are $E=2u(r_4)N_s/5$ and $N=N_s/5$, respectively, whereas in the c-ret crystal $E=(u(r_3)+2u(r_4))N_s/4$ and $N=N_s/4$. By comparing the grand potentials, we see that the square crystal overcomes in stability the c-ret crystal for chemical potentials $\mu$ lower than $5u(r_3)+2u(r_4)=4.1\epsilon$, corresponding to a reduced pressure $Pa^2/\epsilon=(\mu-2u(r_4))/(5\epsilon)=1.3$. A similar comparison between the square crystal and the $T=0$ vapor (i.e., the empty lattice) sets the corresponding transition at $\mu=2u(r_4)=-2.4\epsilon$, or $P=0$.

To sketch the complete phase diagram of the MOVB model we use the transfer-matrix method (see e.g. \cite{runnels1}), which computes the exact pressure as a function of $T$ and $\mu$ for the system defined on a lattice ``strip'' $L\times\infty$, being finite in the row direction and infinite in the other (with periodic conditions at the boundaries of a row). Due to the rather long range of interaction, the basic lattice unit for the definition of the transfer matrix consists of {\em two} rows, implying that the size of the transfer matrix equates the number of states of $2L$ sites. In turn, the pressure is given in terms of the dominant eigenvalue $\lambda_1$ of the transfer matrix as
\be
P=\frac{1}{2L}k_{\rm B}T\ln\lambda_1\,,
\label{eq2}
\ee
where $k_{\rm B}$ is the Boltzmann constant. The quantity $\lambda_1$ in (\ref{eq2}) should be evaluated numerically. In practice, we can take advantage of a few symmetries to reduce the size of the matrix while keeping the maximum eigenvalue unchanged (see Appendix A). Once $P$ has been determined, the number density $\rho$ and the isothermal compressibility $K_T$ are obtained from the formulae
\be
\rho=\left.\frac{\partial P}{\partial\mu}\right|_T\,\,\,\,\,\,{\rm and}\,\,\,\,\,\,K_T=\frac{1}{\rho^2}\left.\frac{\partial\rho}{\partial\mu}\right|_T\,.
\label{eq3}
\ee
Peaks of $\partial\rho/\partial\mu$ or $K_T$ as a function of $\mu$ at fixed $T$ may be taken as indication of singularities in the thermodynamic limit (see Section 3.1), but only if the peak height scales as a specific power of $L$.

In order that both crystalline phases of the MOVB model fit into the lattice, $L$ should be a multiple of 10. It turns out that the only viable case is $L=10$, since for $L=20$ the transfer matrix is huge. For the OVB model $L$ should be a multiple of 5, and we are able to treat lattice strips with up to 20 sites in a row.

To assess, and where necessary, strengthen our transfer-matrix predictions, we carry out grand-canonical Monte Carlo (MC) simulations of the MOVB model (Section 3.2) with single-site moves: at each MC step, we attempt to flip the occupancy of a randomly chosen site; then, the move is accepted or rejected in accordance with the Metropolis criterion. Each simulation run starts from a typical equilibrium configuration of the system at a nearby state point; after a long equilibration, we generate a trajectory a few million cycles long --- one MC cycle consisting of $N_s$ trial moves. By dividing the production run in large blocks, statistical errors are estimated as sample standard deviations of block averages. In addition to density and energy, we compute the isothermal compressibility through the fluctuations of particle number, according to the well-known formula
\be
\rho k_{\rm B}TK_T=\frac{\langle(\delta{\cal N})^2\rangle}{\langle{\cal N}\rangle}\,,
\label{eq4}
\ee
where ${\cal N}=\sum_ic_i$ is the current particle number, $\langle\cdots\rangle$ denotes a grand-canonical average, $\rho a^2=\langle{\cal N}\rangle/N_s$ is the reduced density, and $\delta{\cal N}={\cal N}-\langle{\cal N}\rangle$. We also monitor the density histogram, $P(\rho)=\langle\delta_{{\cal N},\rho V}\rangle$ (denoting $\delta$ the Kronecker delta and $V=N_sa^2$ the system volume), which informs on the ``strength'' of any transformation involving a distinct density change.

The last methodology implemented is umbrella sampling (US), which we apply to the determination of the nucleation barrier for the liquid-to-(square) crystal transition of the MOVB model, so as to confirm that the onset of solid occurs by the same process of thermal activation that works in the continuum (Section 3.3). An important test will be to show that the cost of solid formation decreases with increasing liquid supersaturation. Aside from a paper on the Potts lattice gas~\cite{duff}, we do not know of any other investigation of crystal nucleation in a realistic lattice system. Sophisticated methods exist for the nucleation barrier of the Ising lattice gas~\cite{schmitz}, which however cannot be easily adapted to the solid-liquid transition of interest here.

Given a criterion to identify solidlike particles within a predominantly liquid system, and choosing the size $n$ of a solid cluster as unique reaction coordinate, the work of cluster formation (namely, the free-energy difference between the supersaturated liquid with and without a solid cluster) reads $\Delta\Omega(n)=-k_{\rm B}T\ln(N_n/N_s)$~\cite{tenWolde,maibaum}, where $N_n$ is the average number of $n$-clusters per configuration (see Appendix B). The maximum of $\Delta\Omega(n)$ is the height of the nucleation barrier, whereas its abscissa is the critical cluster size, discriminating (in probabilistic terms) between extinction and growth. However, for low to moderate supercooling/overcompression, the spontaneous occurrence of a large solid cluster in the metastable liquid is a rare event. This poses a problem of poor statistics in the estimate of $N_n$ by MC, which in a US simulation is overcome by the use of a biasing potential that couples with the size $n_{\rm max}$ of the largest cluster. By properly re-weighting the sampled microstates, one eventually recovers the ordinary ensemble averages (more details on the technicalities of the US method in \cite{prestipino3}). In the present study, the biasing potential is 0 in a window around the target size, while being infinite otherwise. The main obstacle to the calculation of $\Delta\Omega(n)$ is the need of identifying the largest cluster after every MC move. This complication can be largely mitigated by the use of a hybrid scheme~\cite{gelb,prestipino4}.

\section{Results}

\subsection{Transfer-matrix phase diagram of the MOVB model}

The transfer-matrix method provides an elegant as well as exact solution to the many-body problem, at least in cases where the transfer matrix is not too big. First considering the OVB model, we extend the transfer-matrix calculations in \cite{orban} to $L=20$, with the purpose to validate the conclusions reached in that paper. We choose two values for the inverse temperature $\beta=(k_{\rm B}T)^{-1}$, i.e., $\eta\equiv e^{\beta\epsilon}=6.5$ and $\eta=9$, the latter pretty close to the triple-point value ($\eta\approx 9.5$). In Fig.\,2 we report the density $\rho$ and the reduced compressibility $\rho k_{\rm B}TK_T$ as a function of $\beta\mu$ for three values of $L$. Of the two peaks in $\rho k_{\rm B}TK_T$, the one for lower $\mu$ refers to the liquid-vapor transition whereas the other peak highlights the transition from liquid to solid. While for both temperatures the freezing transition is very sharp, the liquid-vapor transition is milder, at least for $\eta=6.5$ where it likely corresponds to a smooth crossover.

\begin{figure}
\begin{center}
\includegraphics[angle=-90,width=12cm]{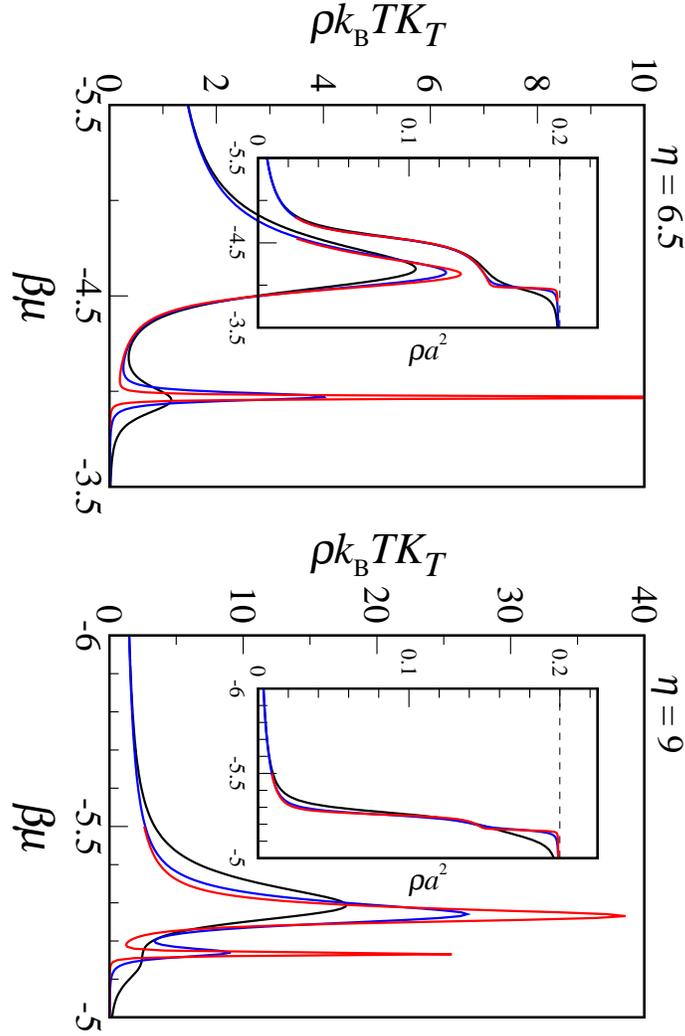}
\end{center}
\caption{Transfer-matrix data for the OVB model at two different temperatures (left panel: $\eta=6.5$; right panel: $\eta=9$) and for three sizes ($L=10$, black; $L=15$, blue; $L=20$, red). For each temperature, the reduced compressibility (main figure) and the density (inset) are plotted as a function of $\beta\mu$. The ideal-gas limit $\rho k_{\rm B}TK_T=1$ is recovered for $\mu\rightarrow -\infty$.}
\end{figure}

\begin{figure}
\begin{center}
\includegraphics[width=12cm]{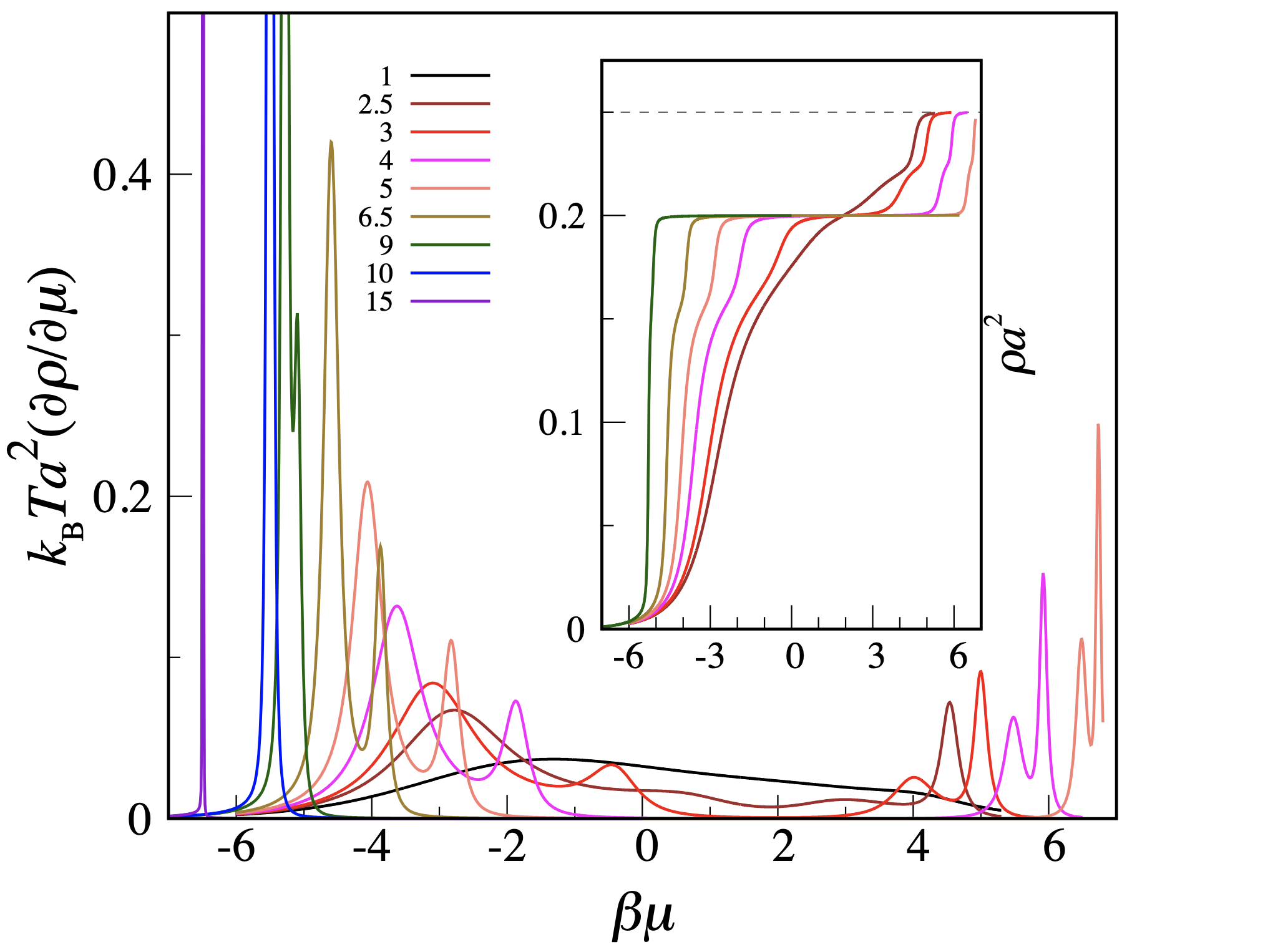}
\end{center}
\caption{Transfer-matrix data for the MOVB model at various temperatures ($\eta$ values are between 1 and 15, see legend). Only results for $L=10$ are available. In the main figure, the $\beta\mu$ derivative of the density is plotted as a function of $\beta\mu$. In the inset, a few density plots are shown. Any peak of the density derivative signals a more or less steep rise in the density, in turn indicative of the possibility of a phase transition in the thermodynamic limit.}
\end{figure}

Moving to the MOVB model, we only have data for $L=10$. In Fig.\,3 we plot the $\beta\mu$ derivative of the density as a function of $\beta\mu$ for a number of $\eta$ values between 1 and 15, along with a number of density profiles (in the inset). We see that, while the low-$\mu$ regime is nearly identical for the OVB and MOVB models, the high-$\mu$ regime is completely different, since a second crystalline phase appears in the MOVB model, heralded by the double-peak structure of the density derivative and confirmed by the density profile. The double (rather than single) peak indicates that the square crystal will not directly transform into the c-ret crystal, but rather through an intermediate fluid phase.

\begin{figure}
\begin{center}
\includegraphics[angle=-90,width=12cm]{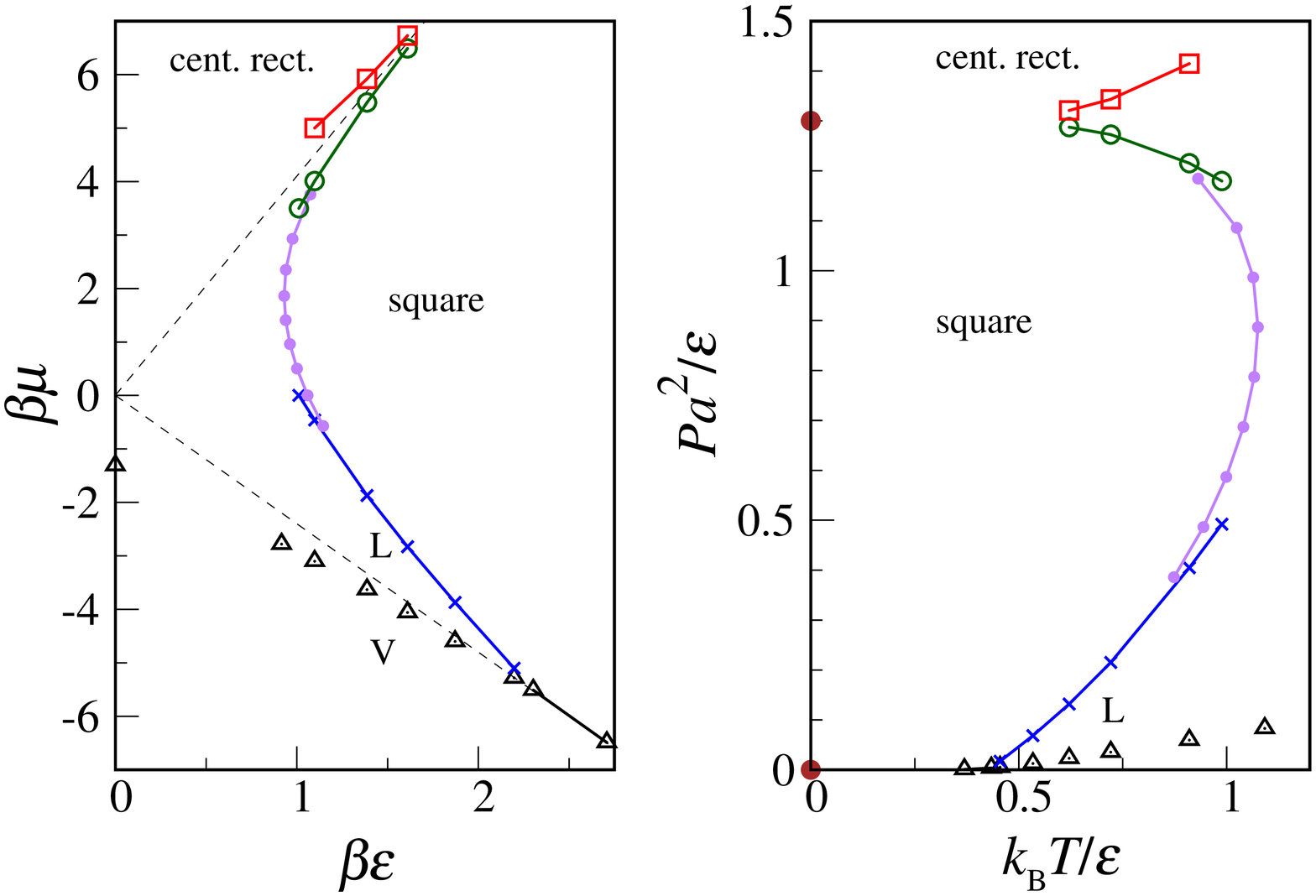}
\end{center}
\caption{MOVB phase diagram according to the transfer-matrix analysis. Left: $\beta\mu$ vs. $\beta\epsilon$. Different types of ``transition points'' are marked with different symbols and colors. The purple dots were computed through scans made at fixed $\mu$ (see text). The two dashed lines represent extrapolations to infinite temperature of the low-$T$ transition loci $\mu=-2.4\epsilon$ (between square crystal and vapor) and $\mu=4.1\epsilon$ (between c-ret crystal and square crystal), see Section 2. Right: $T$-$P$ phase diagram. The brown dots are the $T=0$ transition pressures computed in Section 2.}
\end{figure}

A neater picture emerges from the phase diagram. Taking the location of peaks in the density derivative as the finite-size estimate of transition points, we obtain the diagrams plotted in Fig.\,4 (only the purple dots in both panels were obtained through scans along constant-$\mu$ loci, see more below). In the low-$\mu$/low-$P$ sector we recognize the same phase diagram of the OVB model, but with larger and larger deviations as $\mu$ progressively increases. Beyond a certain $\beta\mu$ value, the solid-liquid locus bends towards low temperatures, implying the existence of a maximum melting temperature for the square crystal. Above $\mu=4.1\epsilon$ (or $P=1.3\epsilon a^{-2}$) the stable solid is the c-ret crystal. With the possible exception of very low $T$, the dense fluid creeps in between the two solids.

\begin{figure}
\begin{center}
\includegraphics[width=12cm]{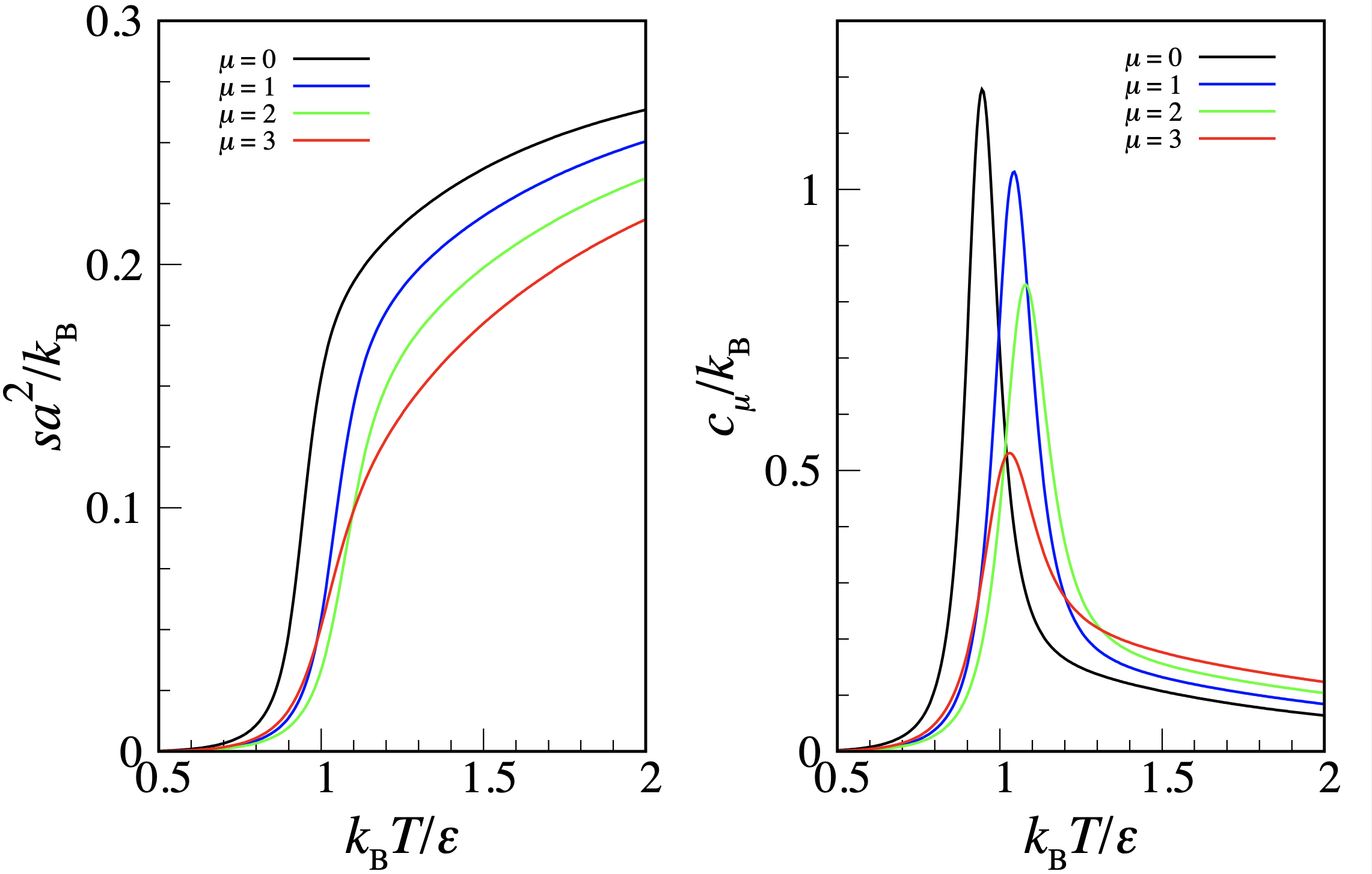}
\end{center}
\caption{Transfer-matrix results for the MOVB model ($L=10$) along a number of constant-$\mu$ lines (in the legend). Left: entropy density; right: constant-$\mu$ specific heat per unit volume.}
\end{figure}

From the knowledge of the $P$ dependence on $T$ for fixed $\mu$ we can derive the entropy density $s$ and the constant-$\mu$ specific heat per unit volume $c_\mu$ by the formulae:
\be
\left.\frac{\partial P}{\partial T}\right|_\mu=s\,\,\,\,\,\,{\rm and}\,\,\,\,\,\,c_\mu=T\left.\frac{\partial s}{\partial T}\right|_\mu\,.
\label{eq5}
\ee
A sample of these quantities is shown in Fig.\,5. The abrupt fall of entropy on cooling occurs at the crossover from fluid to square crystal. Hence, the peak in the $T$-derivative of the entropy (or, alternatively, the specific-heat maximum) provides an estimate of the square crystal-to-fluid transition point. The imperfect matching in Fig.\,4 between the ``coexistence loci'' obtained from $\mu$ and $T$ scans is a finite-size effect, all the more evident when peaks are not sufficiently sharp.

To sum up, the transfer-matrix treatment of the $L=10$ strip is sufficient to sketch the complete phase diagram of the MOVB model. Obviously, this analysis is by no means exhaustive, since the low-temperature regime cannot be accessed for such a small $L$. Furthermore, nothing can be said about the exact location of the liquid-vapor critical point.

\subsection{The liquid-vapor transition is in fact a crossover}

A long-standing issue in statistical physics is how to distinguish between first- and second-order phase transitions in a finite system. At a first-order transition, the specific heat and the compressibility exhibit delta-function singularities in the thermodynamic limit. This should be contrasted with a second-order transition, where the second-order free-energy derivatives diverge algebraically. An infinite-volume system does not anticipate a first-order transition as the transition point is approached. Finite-volume systems do instead anticipate the onset of a phase transition of any order. This feature is exploited by numerical methods, which examine the finite-size scaling (FSS) of extrema of quantities being singular in the thermodynamic limit at the transition point. In finite systems the counterpart of these singularities are smooth peaks, the height and shape of which depend on the strength of the phase transition.

According to the theory of FSS~\cite{fisher1,fisher2,fernandes}, at a first-order transition the height of the compressibility peak on a symmetric lattice increases linearly with the volume $V$ and its width at half maximum shrinks as $1/V$; at a second-order transition, the peak has a slower increase in height $\sim V^{\gamma/(d\nu)}$ (with $\gamma/(d\nu)<1$), and a broader width $\sim V^{-1/(d\nu)}$ ($1/(d\nu)<1$), where $\gamma$ and $\nu$ are the usual critical exponents and $d$ is the dimensionality of space.

In the light of the above arguments, we reconsider the transfer-matrix evidence for condensation and freezing in the MOVB model. A good example is provided by the data reported in Fig.\,2. Though actually referring to the OVB model, these data would also apply for the MOVB model --- in view of the relatively low $\mu$ values. The easier case is freezing, where the delta-function-type increase of compressibility is the clear imprint of a first-order transition. Less clear is the status of the liquid-vapor transition, since both the increase of the peak and the reduction in width are rather slow, even near the triple point. To settle the question, we have carried out Monte Carlo simulations of the OVB and MOVB models for $\eta=9$, considering $L\times L$ lattices of various sizes (up to $L=120$), in order to see how the compressibility behaves near the transition from vapor to liquid as a function of $L^2$. Equilibrium averages are computed over two million MC cycles. Our results, reported in Fig.\,6, clearly indicate that the compressibility converges to a finite value with increasing $L$, meaning that no strict phase transition actually occurs here. A similar conclusion follows from the plot of the density distribution (see Fig.\,7, which refers to $L=120$). We observe a smooth changeover from a vapor-like peak to a liquid-like peak, passing through a broad histogram for $\mu=-5.27$ with no evidence for a valley between peaks, which makes us envisage the absence of a free-energy barrier also in the infinite-volume limit.

To conclude, in the (M)OVB model liquid and vapor are only loosely separated (i.e., there is no sharp distinction between the two) even in the thermodynamic limit. This may come as a surprise, considering that even the simple Ising lattice gas has a liquid-vapor transition. However, when the range of exclusion extends beyond the central particle, the lattice-gas model actually represents a fluid-solid model. In this case we expect the onset of sublattice order at low temperature/high pressure, which would generally be accompanied by a first-order transition (due to the symmetry breaking involved). Whether the inclusion of an attraction outside the core induces a further liquid-vapor transition probably depends on the symmetry and dimensionality of the underlying lattice and the range of the attraction.

\begin{figure}
\begin{center}
\includegraphics[angle=-90,width=12cm]{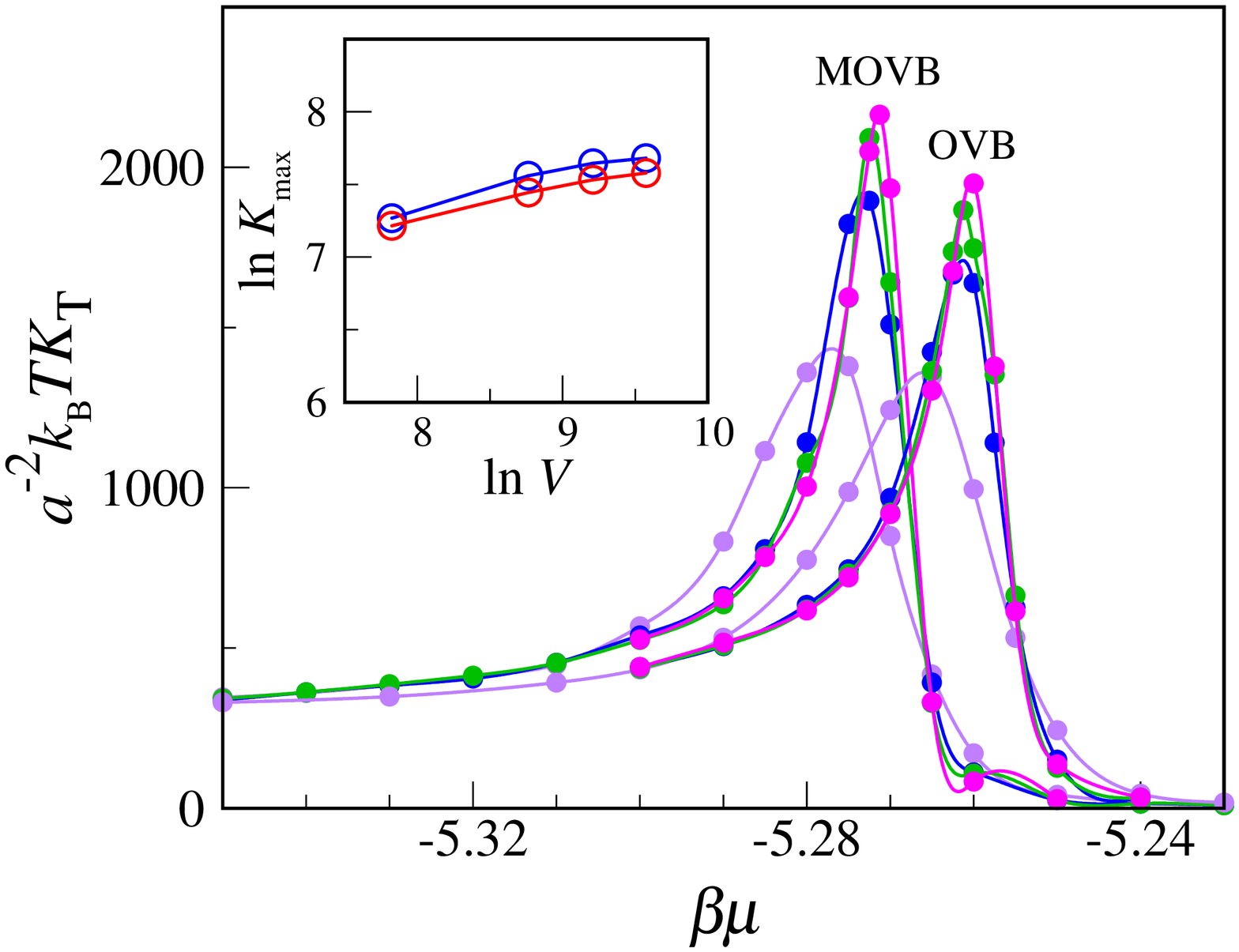}
\end{center}
\caption{Compressibility data for the OVB and MOVB models near the supposed liquid-vapor transition point for $\eta=9$. We have considered square lattices of four different lateral sizes $L$: 50 (purple), 80 (blue), 100 (green), and 120 (magenta). The smooth lines through the data points are spline interpolants. The statistical uncertainties are negligible, i.e., smaller than the size of the symbols. In the inset, the maximum of $k_{\rm B}TK_T$ is reported vs. volume on a log-log scale for both models, to show that in the thermodynamic limit no phase transition is likely to occur in either of the models.}
\end{figure}

\begin{figure}
\begin{center}
\includegraphics[width=12cm]{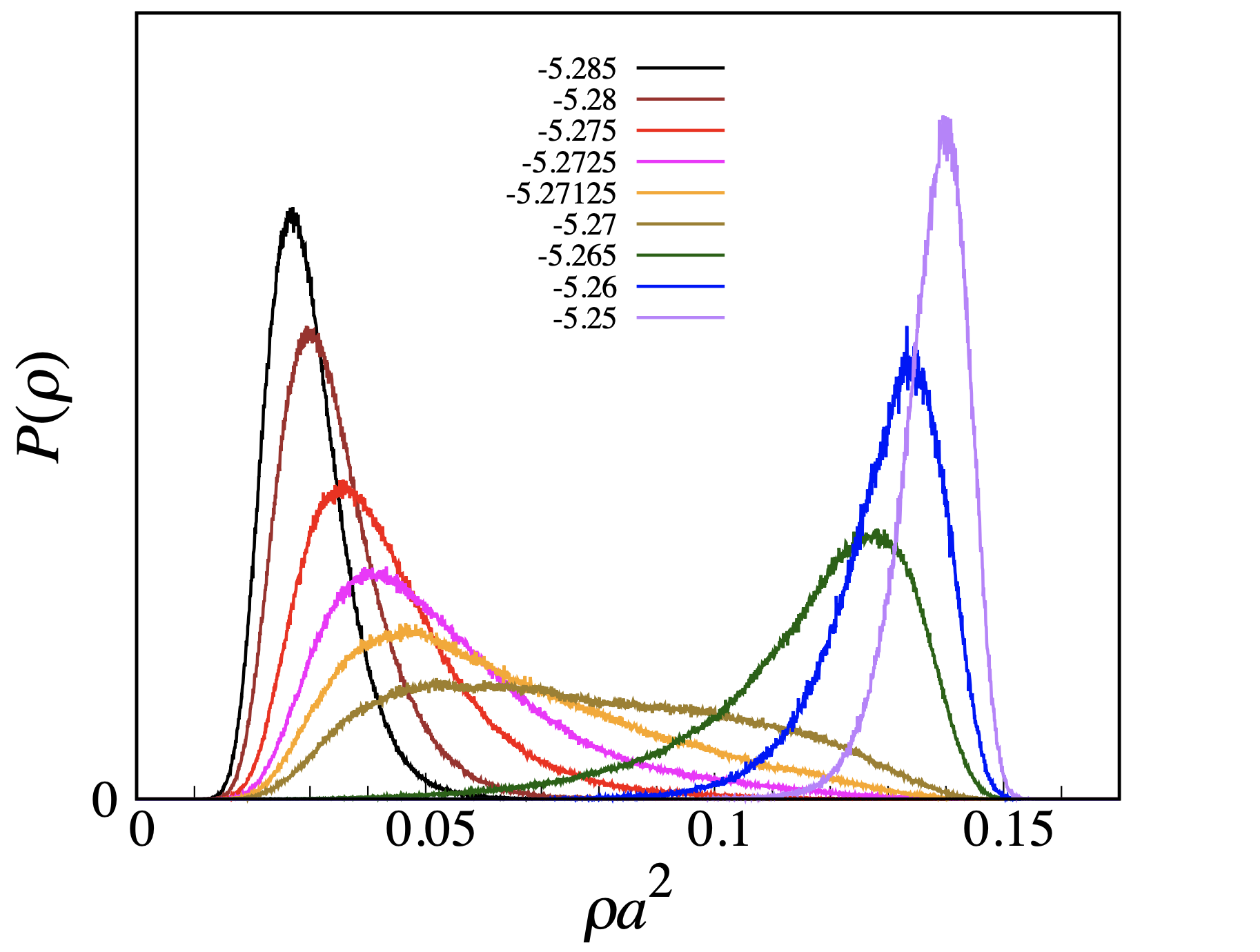}
\end{center}
\caption{MOVB model for $\eta=9$ and $L=120$. Probability distribution of the density across the apparent liquid-vapor transition ($\beta\mu$ values in the legend).}
\end{figure}

\subsection{Features of crystal nucleation from the ``liquid''}

Over the last decades, there has been considerable interest in the study of crystal nucleation from the liquid~\cite{sear,sosso}. Describing this process in detail is of great importance for many practical applications (such as drug synthesis and production, ice formation in clouds, and prevention of amyloid diseases), all impacted by the issue of polymorph selection during the early stages of solidification. However, nucleation is of utmost relevance primarily for fundamental reasons, since any discontinuous phase transition is triggered by the formation of a droplet of the stable phase (at least, up to precursors~\cite{russo}). Here we exploit the universality of this connection as a key to demonstrate that the freezing transition in the MOVB model is of standard type.

Besides the evidence provided by the compressibility, the first-order character of freezing is also evident in the hysteresis found in sequences of MC runs initiated from either side of the transition point. For instance, for $\eta=6.5$ we have been able to overcompress the MOVB liquid up to $\beta\mu=-3.65$ (well above the nominal transition point at $\beta\mu\simeq -3.87$) and to expand the solid down to $\beta\mu=-4.15$. Not surprisingly, we instead found no little trace of hysteresis across the liquid-vapor ``transition''. To characterize the resistance of the metastable liquid to conversion into solid, we calculate the free-energy cost of solid-cluster formation as a function of cluster size, using $\mu$ as a driving parameter (no attempt will be made to estimate the nucleation rate).

Preliminary to any study of crystal nucleation is the choice of a local measure of crystallinity, which could enable the distinction, in any system configuration, between solidlike and liquilike particles. To simplify things, we consider a metastable liquid at moderate pressure, so as to avoid the competition between different polymorphs (in the reasonable assumption that, in the relevant range of densities, there would be no crystalline structure capable to compete in energy with the square crystal). In this case, the crystallinity criterion can be directly tuned to the target solid structure (that is, to the square crystal).

Consider a configuration of the metastable liquid. We attach the ``solidlike'' label to any particle forming bonds with two or more particles at distance $r_4$ from it, but only if at least two of these bonds are perpendicular to each other. Let 1 be a solidlike particle forming two mutually perpendicular bonds with particles 2 and 3; then, the triplet $\{1,2,3\}$ is called a ``wedge of center 1'' (we may also say that particle 1 is solidlike in the given configuration if it is the center of a wedge). Two solidlike particles, say 1 and 2, will be part of the same cluster if 1 belongs to a wedge of center 2, and {\em vice versa}. With these rules established, we can i) identify the solidlike particles present in the given configuration and ii) enumerate the connected assemblies (clusters) of solidlike particles by the Hoshen-Kopelman algorithm~\cite{hoshen}. A typical outcome of our clustering algorithm is illustrated in Fig.\,8, which refers to a metastable liquid for $\eta=6.5$ and $\beta\mu=-3.70$. We see that most solid clusters, even the smallest ones, have a distinct square-ordered structure. Occasionally, we see two particles at $r_3$ distance within the same cluster, which in this case can be considered as ``polycrystalline''. Clearly, we could have dubbed solidlike any particle at the center of two or four wedges or might have chosen a different descriptor of local order, such as a Steinhardt order parameter~\cite{steinhardt} or the like~\cite{lechner}, but in this case the only difference would be in the statistics of small clusters and in the critical size, with little influence on the height of the nucleation barrier~\cite{filion,prestipino3}.

\begin{figure}
\begin{center}
\includegraphics[width=12cm]{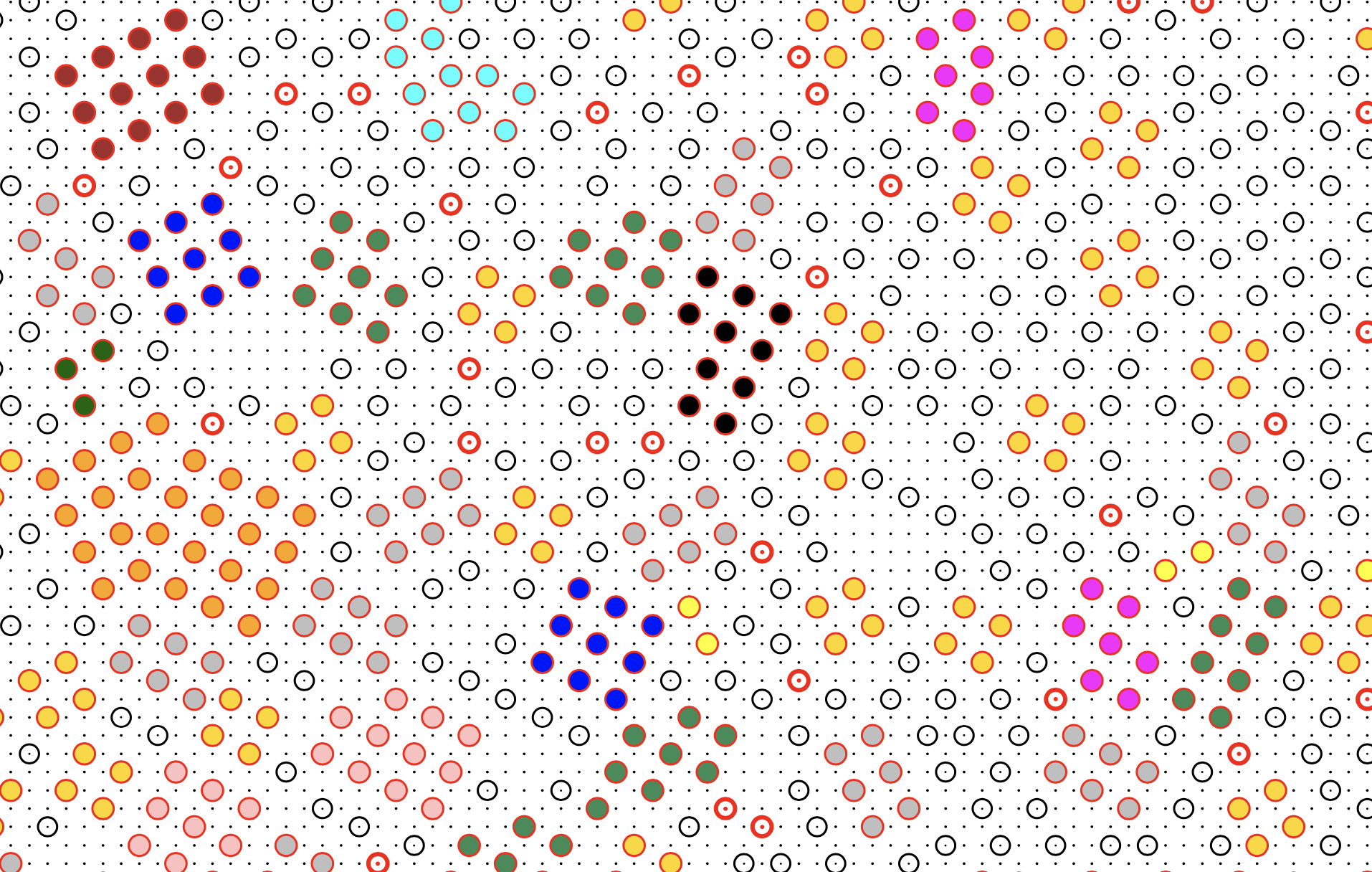}
\end{center}
\caption{MOVB model on a $100\times 100$ lattice, for $\eta=6.5$ and $\beta\mu=-3.70$: typical configuration (particular) of the overcompressed liquid, with liquidlike (white circles) and solidlike particles (colored circles) well distinguished. For the present values of $\eta$ and $\beta\mu$ the reduced density is about $0.161$ and the energy per particle is $-0.201\epsilon$. The size of the maximum cluster in the configuration shown is 28. Different colors are used to represent particles belonging to solid clusters with different sizes. The white circles with the red contour are isolated solidlike particles. The dense grid in the background is the underlying square lattice. Notice the presence of clusters where the occurrence of two particles at distance $r_3$ apart induces a change in crystalline orientation.}
\end{figure}

\begin{figure}
\begin{center}
\includegraphics[width=12cm]{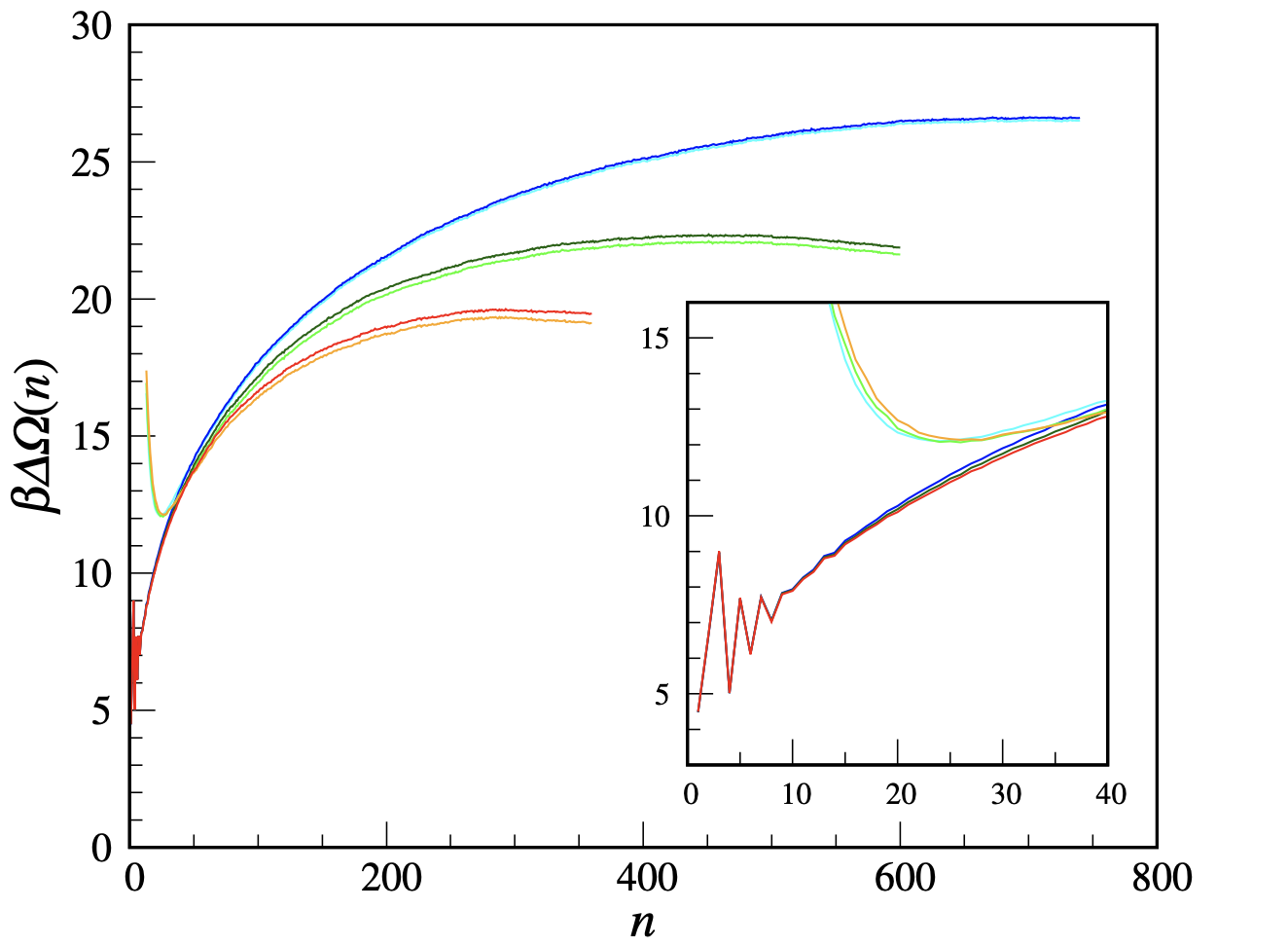}
\end{center}
\caption{MOVB model for $\eta=6.5$ and $\beta\mu=-3.75,-3.70,-3.65$ (from top to bottom): the (reduced) cluster free energy (blue, dark green, and red) is plotted together with the (reduced) cost of formation of the largest cluster shifted upwards by $\ln N_s$ (cyan, light green, and orange). In the inset, a magnification of the low-size region.}
\end{figure}

We are now in a position to compute the free-energy cost of a solid cluster, using the US method. We take $\eta=6.5$ and a $100\times 100$ lattice. We divide the $n_{\rm max}$ range in windows, i.e., $[n_0-10,n_0+10]$ with $n_0=10,20,\ldots$, and carry out the simulations in sequence, performing $10^5$ cycles for each $n_0$ to equilibrate the system, followed by a five times longer production run (which proved sufficient to determine the nucleation barrier with enough accuracy). Then, the separate free-energy branches are vertically shifted so as to match with each other and with the cluster free-energy curve resulting from an unbiased MC simulation of the metastable liquid (in the latter simulation only the statistics of clusters with size not larger than $\approx 15$ turns out accurate). The final cluster free energy is shown in Fig.\,9 for three supersaturations ($\beta\mu=-3.75,-3.70,-3.65$). For each $\beta\mu$ two quantities are plotted, namely $\beta\Delta\Omega(n)$ and $\beta\Delta\Omega^*(n)+\ln N_s$ (the latter one being defined in Appendix B), which should coalesce for large $n$. The residual discrepancy may be ascribed to statistical uncertainties. All in all, the cluster free energy has the usual shape and dependence on the supersaturation. Only the small-$n$ behavior is non-standard, being strongly $n$ dependent and in the same terms non-monotonic for all $\beta\mu$. This behavior certainly reflects the peculiar definition of crystallinity adopted, which e.g. discourages 3-clusters relative to 2- and 4-clusters. It is by the way clear that the cluster free energy is, by its very definition, a meaningful concept only for large clusters. For large $n$, most solidlike particles are gathered in a single big cluster, just for entropic reasons~\cite{tenWolde2}. The typical shape of the critical cluster for $\beta\mu=-3.70$ can be appreciated in Fig.\,10. Rather than circular, the critical cluster is slightly elongated and has an irregular contour, confirming a non-trivial role in nucleation for the length/area of the cluster boundary~\cite{prestipino5}.

\begin{figure}
\begin{center}
\includegraphics[width=14cm]{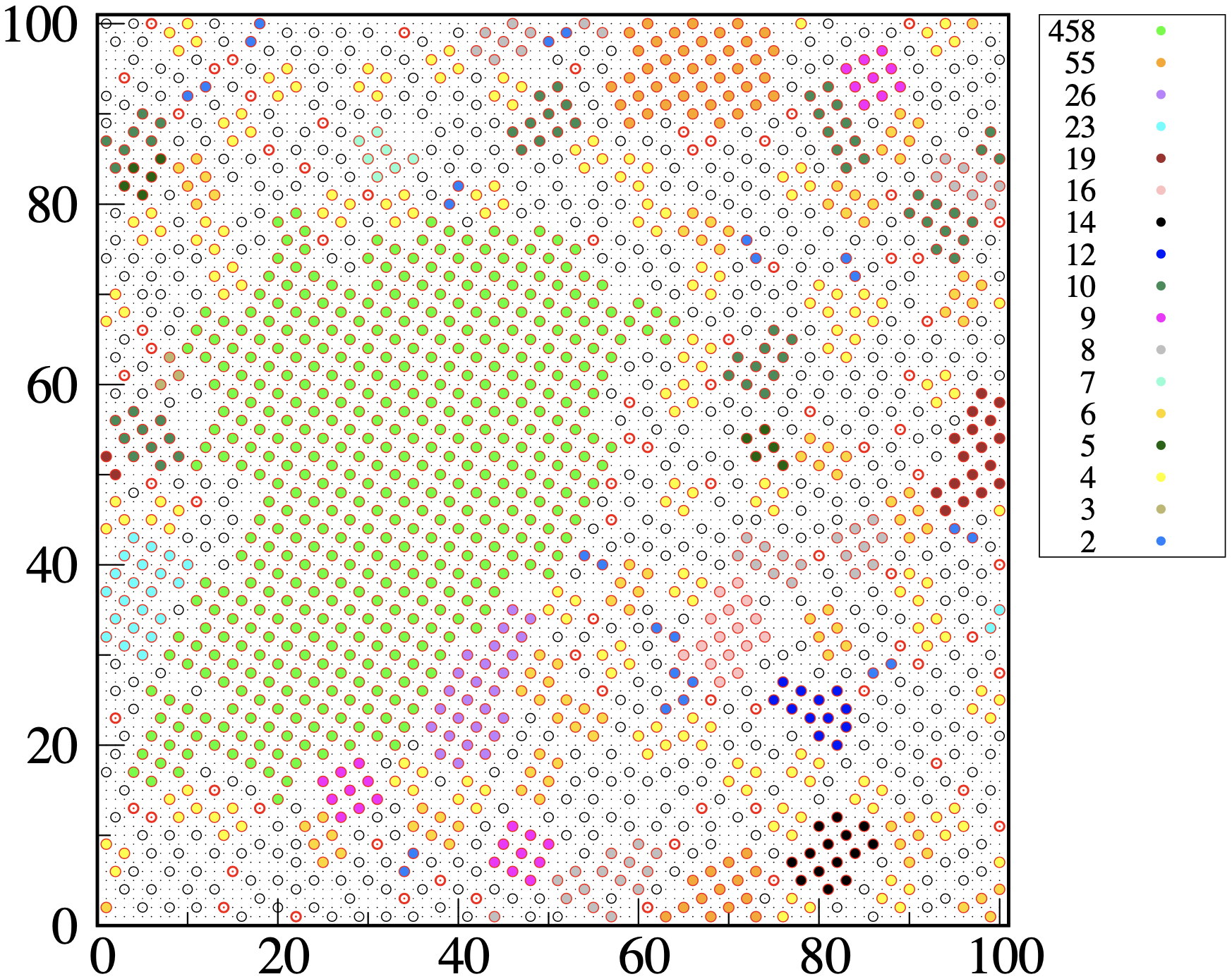}
\end{center}
\caption{MOVB model for $\eta=6.5$ and $\beta\mu=-3.70$: typical configuration of the system near the top of the nucleation barrier. Cluster sizes and colors in the legend.}
\end{figure}

We add a final comment on the large-$n$ fate of the curves in Fig.\,9. For $\beta\mu=-3.75$, we had to stop the US simulation when the extension of the biggest cluster reached the box edge, so as to avoid any spurious influence from the interaction of the cluster with its periodic images. Instead, the US simulation for $\beta\mu=-3.65$ was stopped when the system completely solidified, which happened abruptly through a jump in the density and the formation of a few coexisting big clusters with 300-350 solidlike particles each.

\section{Conclusions}

The phase behavior of lattice-gas systems has both similarities and differences with that of particles in the continuum. An enlightening example is provided by the OVB model~\cite{orban}, a lattice-gas model defined on the square lattice. In \cite{orban} this system was claimed to have the same phase diagram of an ordinary simple fluid, with a square crystal as solid phase. However, this conclusion ensued from a transfer-matrix treatment of a strip only 10 sites wide. In the analysis presented here, we extend this width to 20, but this is still insufficient to assess the nature of the transition from vapor to liquid. We have thus carried out Monte Carlo simulations of $L\times L$ lattices, with $L$ up to 120, by which we definitely exclude that vapor and liquid are distinct phases: their apparent coexistence locus is no more than a disorder line. The conclusion remains if we take the repulsion at third-neighbor lattice distance to be finite, this way switching from the OVB to the MOVB model. However, the latter model at least allows for the existence of a second crystalline phase at high pressure, which brings the MOVB model closer to real-life materials.

Next, we have inquired into the first stages of the transition of ``liquid'' into square crystal, asking whether the nucleation process would occur in the same terms as in continuous three-dimensional systems. To this aim, we introduce a notion of crystallinity tuned to the solid at hand. By computing the nucleation barrier, we find that the cluster free energy as a function of cluster size has the same shape as for ordinary supercooled liquids, except for the smallest sizes where lattice peculiarities cause a characteristic non-monotonic variation.

Our study raises the interesting question as to whether a sharp distinction between liquid and vapor in a lattice-gas model could be achieved by either increasing the range of interaction or changing the underlying lattice. This gives us the opportunity to revisit the conclusions reached in \cite{prestipino1} for a number of triangular-lattice-gas models, for which a more careful analysis of liquid-vapor equilibrium is planned for the next future. Other directions of research development may concern the clustering of two-dimensional lattice particles with overlapping cores (i.e., the discrete-space counterpart of the study in \cite{prestipino6}) or the finite-size phases of particles living on the nodes of a dense polyhedral mesh (much denser than considered in \cite{degregorio}).


\appendix
\section{On the dominant eigenvalue of the transfer matrix}

The calculation of the dominant eigenvalue of the transfer matrix $T^i_j$ can be done numerically by power iteration~\cite{blum}. However, the size of the transfer matrix increases exponentially with the length $L$ of a row, eventually leading to an exceedingly long cpu-time --- if not even to a problem of memory overflow. To partially overcome this limitation, Runnels and Combs have proposed~\cite{runnels2,runnels3} to make use of symmetry properties in order to replace the transfer matrix $T^i_j$ with a smaller matrix $\tau^\alpha_\beta$ {\em without affecting the dominant eigenvalue}. The idea, which we here reproduce for the reader's convenience, is to gather together in the same equivalence class $\alpha$ all the two-row states obtained from a given state $i$ by a horizontal shift (that is, a cyclic permutation of the occupancies) or a reflection relative to the vertical axis of the strip. Then, the matrix
\be
\tau^\alpha_\beta=\sum_{j\in\beta}T^i_j
\label{a-1}
\ee
will not depend on the particular state $i\in\alpha$. Now, by noting that $\tau^\alpha_\beta$ is a primitive matrix~\cite{runnels2,meyer}, it is easy to prove that any eigenvalue $\lambda$ of $\tau^\alpha_\beta$ is also an eigenvalue of $T^i_j$ (while the opposite is false). Let a $u$ vector be given such that
\be
\sum_\alpha\tau^\alpha_\beta u^\beta=\lambda u^\alpha
\label{a-2}
\ee
and take $v^i=u^\alpha$ for all $i\in\alpha$. Then, we obtain:
\be
\sum_jT^i_jv^j=\sum_\beta\left(\sum_{j\in\beta}T^i_jv^j\right)=\sum_\beta\left(\sum_{j\in\beta}T^i_ju^\beta\right)=\sum_\beta u^\beta\left(\sum_{j\in\beta}T^i_j\right)=\sum_\alpha\tau^\alpha_\beta u^\beta=\lambda u^\alpha=\lambda v^i\,.
\label{a-3}
\ee
Therefore, the eigenvalue $\lambda$ of $\tau^\alpha_\beta$ is also an eigenvalue of $T^i_j$ and the corresponding eigenvector is $v$. Since all the components of the dominant eigenvector of $\tau^\alpha_\beta$ are positive (by the Perron-Frobenius theorem), also the corresponding $v$ vector has positive components, hence it will be the dominant eigenvector of $T^i_j$ (again, by the Perron-Frobenius theorem). This implies that the dominant eigenvalue of $\tau^\alpha_\beta$ coincides with the dominant eigenvalue of $T^i_j$.

The number $m(L)$ of equivalence classes (i.e., the size of $\tau^\alpha_\beta$) is typically much smaller than the original number $n(L)$ of two-row states. For the MOVB model, $n(10)=1025$ and $m(10)=78$. However, for $L=20$ the matrix $\tau^\alpha_\beta$ is already so huge ($n(20)=1048577$ and $m(20)=27012$) that we could not manage to store it in the memory of our computer. We have less problems with the OVB model where, due to a more extended core, the sizes of $T^i_j$ and $\tau^\alpha_\beta$ are smaller. For example, $n(20)=196333$ and $m(20)=5140$ (which is still amenable).

\section{Cluster free energy and its relation to the cluster-size distribution}
\setcounter{equation}{0}
\renewcommand{\theequation}{B\arabic{equation}}

There is a simple expression for the free energy cost $\Delta\Omega(n)$ of a $n$-sized cluster in terms of the cluster-size distribution. Assume that we have preliminarily identified, by some reasonable criterion, the solid-like particles present in the given configuration ${\bf c}=\{c_1,\ldots,c_{N_s}\}$. Following Maibaum~\cite{maibaum}, let $s_i({\bf c})$ be the size of the cluster containing the $i$th site (under the assumption that $c_i=1$ and this particle is solidlike), whereas $s_i({\bf c})=0$ otherwise. Then, the number $N_n({\bf c})$ of $n$-clusters reads:
\be
N_n({\bf c})=\frac{\sum_{i=1}^{N_s}\delta_{s_i({\bf c}),n}}{n}\,\,\,\,\,\,(n\ge 1)\,,
\label{b-1}
\ee
and its thermal average is:
\be
{\cal N}_n=N_s\frac{\left\langle\delta_{s_1,n}\right\rangle}{n}\,.
\label{b-2}
\ee
Assuming that there is a cluster containing site 1, the probability that it has size $n$ is
\be
P(s_1=n\cap\,{\rm there}\,\,{\rm is}\,\,{\rm a}\,\,{\rm solidlike}\,\,{\rm particle}\,\,{\rm in}\,\,1)=\left\langle\delta_{s_1,n}\right\rangle\,.
\label{b-3}
\ee
Then, the cluster free energy (in reduced, $k_{\rm B}T$ units) reads
\be
\beta\Delta\Omega(n)\equiv-\ln P(s_1=n\cap\,{\rm there}\,\,{\rm is}\,\,{\rm a}\,\,{\rm solidlike}\,\,{\rm particle}\,\,{\rm in}\,\,1)+\ln n=-\ln\frac{{\cal N}_n}{N_s}\,,
\label{b-4}
\ee
by taking into account the degeneracy implicit in the choice of a particular cluster particle.

When $n$ is large enough, all solidlike particles belong to a single cluster, which is also the largest cluster: $s_i({\bf c})=S({\bf c})$ for each site $i$ in the cluster (denoting with $S({\bf c})$ the size of the largest cluster in {\bf c}). Hence, $\sum_{i=1}^{N_s}\delta_{s_i({\bf c}),n}=S({\bf c})\delta_{S({\bf c}),n}=n\delta_{S({\bf c}),n}$ and $N_n({\bf c})=\delta_{S({\bf c}),n}$. As a result,
\be
\beta\Delta\Omega(n)=\beta\Delta\Omega^*(n)+\ln N_s\,,
\label{b-5}
\ee
where
\be
\beta\Delta\Omega^*(n)=-\ln\left\langle\delta_{S({\bf c}),n}\right\rangle
\label{b-6}
\ee
represents the reduced free energy associated with the probability distribution of $S({\bf c})$. For small $n$ the two sides of Eq.\,(\ref{b-5}) yield different numbers and, as argued by Maibaum~\cite{maibaum}, the r.h.s. may be expected to be larger than the l.h.s., due to the penalty involved in constraining $S({\bf c})$ to a value smaller than its typical value in the metastable liquid.


\end{document}